\documentclass[pre,twocolumn]{revtex4}
\usepackage{epsfig}
\usepackage{bm}
\usepackage{amssymb}
\usepackage{amsmath}

\begin{document}

\title{Lagrangian chaos and turbulence in fluid dynamics}

\author{A. Bershadskii}

\affiliation{
ICAR, P.O. Box 31155, Jerusalem 91000, Israel
}

\begin{abstract}

  Randomization of the Lagrangian chaos in fluid dynamics has been analyzed using results of direct numerical simulations, laboratory measurements, and oceanic observations. The notion of distributed chaos has been used in order to quantify this phenomenon (the main dimensionless parameter of distributed chaos has been used as a measure of randomization). The analysis includes isotropic homogeneous fluid motion, buoyancy-driven flows, particle-laden flows, magnetohydrodynamic flows, and subsurface oceanic mixing. The role of the dissipative Birkhoff-Saffman and Loitsyanskii invariants as well as the invariants related to the spontaneous breaking of the local reflectional symmetry (in particular, the Levich-Tsinober invariant) in the randomization phenomenon has been also investigated.

\end{abstract}

\maketitle

\section{Introduction}

  Lagrangian approach can be very useful in studying mixing and dispersion in fluid dynamics. In a Lagrangian approach, an observer follows the motion (the trajectories time evolution) of fluid particles, unlike an Eulerian approach where an observer considers the velocity field ${\bf u}({\bf x}, t)$ in a space domain at any time $t$. It is known that even a regular velocity field can produce rather irregular motion of the corresponding fluid particles (see, for instance, paper Ref. \cite{aref}). Sensitivity to the initial conditions (the exponential divergence of the initially nearby trajectories of the fluid particles) is the main characteristic of the Lagrangian chaos.\\
  
    The notion of smoothness can be used as a basis for classification of the chaotic regimes. Smooth dynamics is generally characterized by  the {\it stretched} exponential Lagrangian frequency spectra
$$
E(\omega) \propto \exp-(\omega/\omega_{\beta})^{\beta}.  \eqno{(1)}
$$     
 where $1 \geq \beta > 0$ and $\omega$ is the frequency. The value $\beta =1$ (i.e. the exponential spectrum):
$$ 
E(\omega) \propto \exp(-\omega/\omega_c),  \eqno{(2)}
$$ 
can be associated with deterministic chaos (see, for instance, Refs. \cite{fm}-\cite{mm2}). 

  For $ \beta < 1$ the Lagrangian chaotic-like dynamics is not deterministic but nevertheless smooth (the distributed chaos, see below). The non-smooth randomized dynamics is usually characterized by the scaling (power-law) spectra (hard turbulence \cite{wu}). \\
  
  In the distributed chaos (i.e. for $\beta <1$) the value of $\beta$ can characterize a measure of randomization. Namely, lesser values of the dimensionless parameter $\beta$ correspond to stronger randomization of the fluid dynamics. One can see in Fig. 1 that smaller values of the Taylor-Reynolds number $Re_{\lambda}$ correspond to larger values of $\beta$ up to the $\beta = 1$ (corresponding to deterministic chaos). It seems rather natural that larger values of $R_{\lambda}$ should generally correspond to stronger randomization. \\

\begin{figure} \vspace{-0.4cm}\centering 
\epsfig{width=.45\textwidth,file=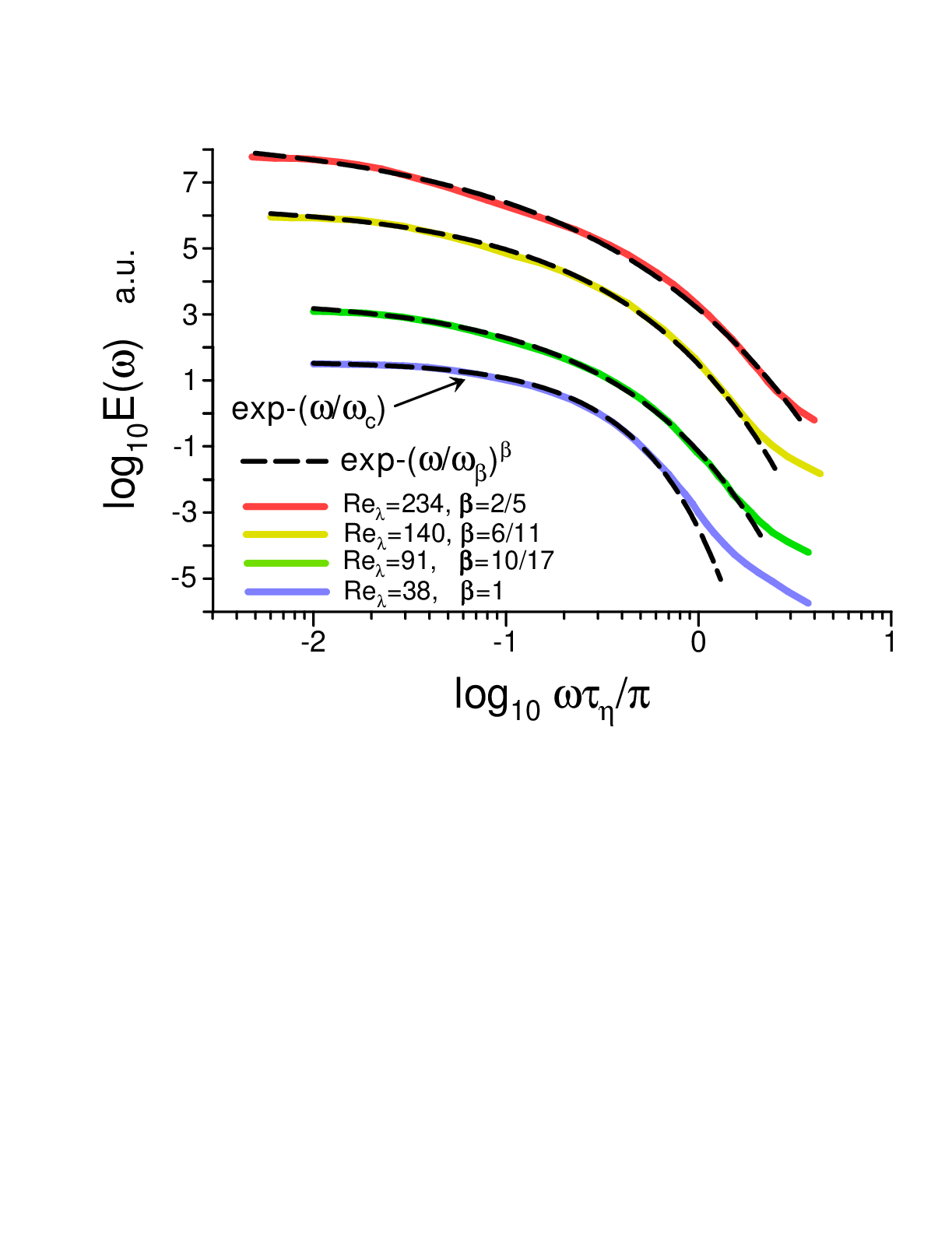} \vspace{-4.4cm}
\caption{Lagrangian frequency spectra of velocity in statistically stationary isotropic homogeneous fluid motion (DNS at different values of $R_{\lambda}$). The spectra are vertically shifted for clarity.} 
\end{figure}
 
    Figure 1 shows the results of direct numerical simulations (DNS) of statistically stationary isotropic homogeneous chaotic/turbulent motion of an incompressible fluid reported in the paper Ref. \cite{pk}. The spectral data were taken from Fig. 5 of the Ref. \cite{pk} ($\tau_{\eta}$ is the Kolmogorov time scale). The DNS was performed for the Navier-Stokes equations. The statistically stationary chaotic/turbulent motion was maintained in the range $38 < Re_{\lambda} <234$ using a large-scale forcing. The dashed curves in Fig. 1 indicate the stretched exponential spectra Eq. (1) and the exponential spectrum Eq. (2) ($\beta =1$, the deterministic chaos for the smallest considered $Re_{\lambda} =38$). \\
    
    In the next sections, the values of $\beta$ shown in the Fig. 1 will be related to different invariants of the fluid dynamics and it will be also shown that spontaneous breaking of the local reflectional symmetry plays an important role in the fluid dynamics (including the isotropic homogeneous motions for sufficiently large $Re_{\lambda}$).  
    
 \section{Lagrangian deterministic chaos in more complex flows}
 
  It was shown in the Introduction that the Lagrangian deterministic chaos (the exponential spectrum) takes place in isotropic homogeneous fluid motion at small $Re_{\lambda}$ (about the Eulerian wavenumber exponential spectra in the isotropic homogeneous fluid motion at small $Re_{\lambda}$ see, for instance, Ref. \cite{kds}).  \\
  
  Let us now give several other examples of the exponential Lagrangian frequency spectra in more complex flows. \\
  
    Figure 2 shows the Lagrangian frequency spectrum of zonal and meridional velocity for trajectories obtained using the subsurface floats at a depth between 400 and 1000 m in the Atlantic Ocean (the World Ocean Experiment - WOCE). The spectral data were taken from Fig. 3 of Ref. \cite{rupolo}. The floats were neutrally buoyant and ballasted for a specific pressure. A total of 132 736 daily observations were used to compute the spectra. \\

\begin{figure} \vspace{-1cm}\centering 
\epsfig{width=.46\textwidth,file=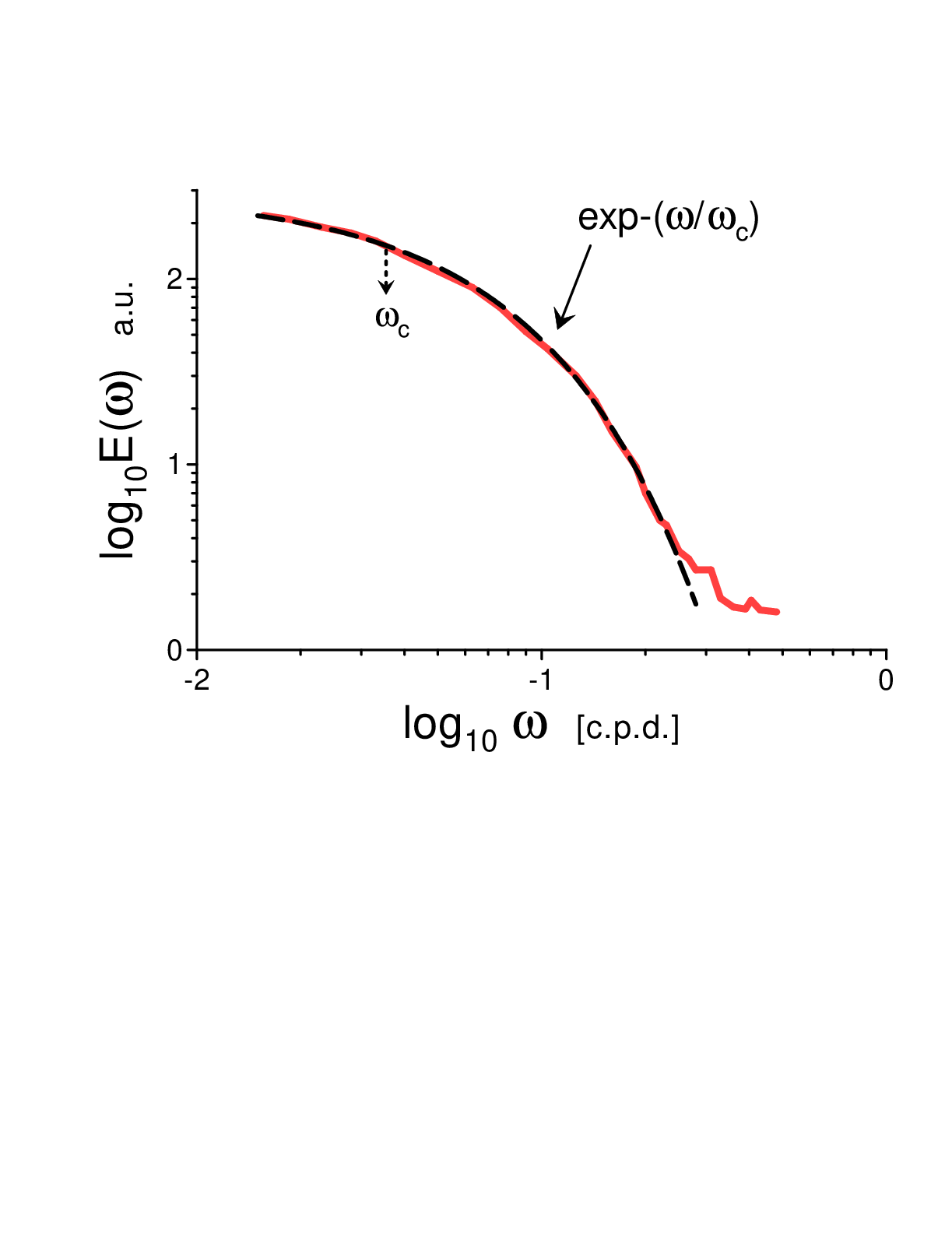} \vspace{-4.4cm}
\caption{Lagrangian frequency spectrum of zonal and meridional velocity for trajectories obtained using the subsurface floats at a depth between 400 and 1000 m in the Atlantic Ocean (the World Ocean Experiment - WOCE).} 
\end{figure}
\begin{figure} \vspace{-0.3cm}\centering
\epsfig{width=.465\textwidth,file=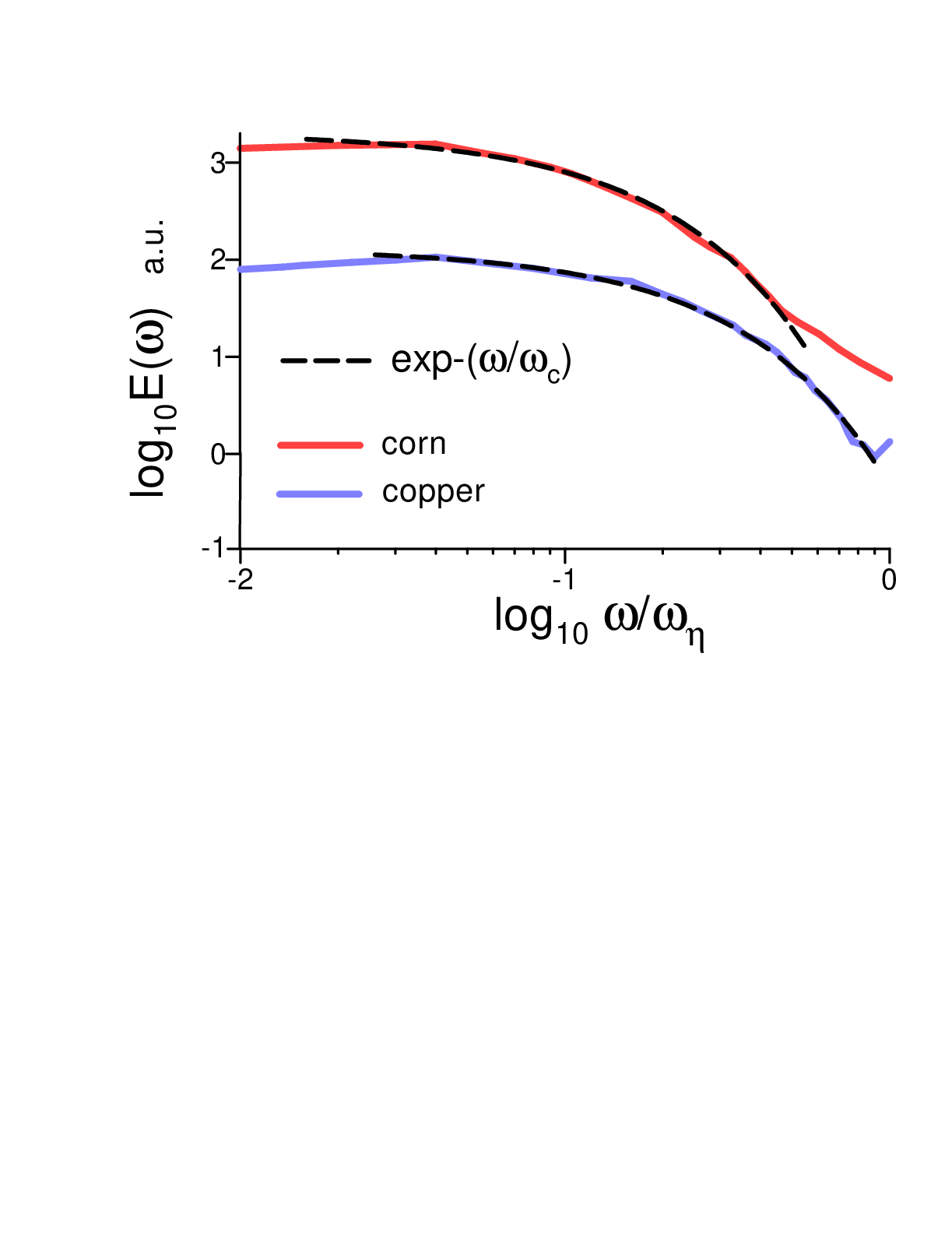} \vspace{-5.24cm}
\caption{The Lagrangian frequency spectra computed for the velocity of fluid surrounding the inertial particles of corn and those of copper for a particle-laden decaying isotropic homogeneous flow in a cubic box. The spectra are vertically shifted for clarity.} 
\end{figure}
   
    The dashed curve in Fig. 2 indicates the exponential spectrum Eq. (2) (the deterministic chaos). The dotted arrow indicates the characteristic frequency $\omega_c$.\\
    
    Figure 3 shows the Lagrangian frequency spectra obtained in a direct numerical simulation (DNS) reported in a paper Ref. \cite{et} $(\omega_{\eta} =\pi/\tau_{\eta}$). The spectral data were taken from Fig. 13b of the Ref. \cite{et}. \\
    
    In this DNS a particle-laden decaying isotropic homogeneous flow in a cubic box (with periodic boundary conditions) was studied using the incompressible Navier-Stokes equations (one-way coupling case) and the complete equation of particle motion (including pressure and viscous drag forces, inertia force of added mass,  force due to viscous stresses and fluid pressure gradient, viscous force related to unsteady relative acceleration, and gravity (or buoyancy) force). The initial $Re_{\lambda}$ was equal to 25 and final  $Re_{\lambda} = 16$.\\ 
    
    The Lagrangian frequency spectra shown in the Fig. 3 were computed for the velocity of fluid surrounding the inertial particles of corn and those of copper (i.e. the local fluid existing around the solid point-like particle at each position of the particle along its trajectory). 
    The dashed curves in Fig. 3 indicate the exponential spectrum Eq. (2) (the deterministic chaos). We will return to this DNS in Section IV. \\

 \section{Lagrangian distributed chaos and dynamic invariants}  
 
  When the characteristic frequency $\omega_c$ in the Eq. (2) randomly fluctuates one needs an ensemble averaging 
$$
E(\omega) \propto \int_0^{\infty} P(\omega_c) \exp -(\omega/\omega_c)~d\omega_c \eqno{(3)}
$$
in order to obtain the average spectrum.

  Providing that the randomized dynamics is still smooth, the spectrum $E(\omega)$ has the stretched exponential form Eq. (1). Comparison of the Eq. (1) and Eq. (3) gives asymptote of  the probability distribution $P(\omega_c)$ for large characteristic frequencies $\omega_c$ \cite{jon}
$$
P(k_c) \propto k_c^{-1 + \beta/[2(1-\beta)]}~\exp(-\gamma k_c^{\beta/(1-\beta)}), \eqno{(4)}
$$     
where $\gamma$ is a constant.\\

 The dissipative dynamics described by the Navier-Stokes equations has two fundamental invariants: Birkhoff-Saffman \cite{bir},\cite{saf},\cite{dav}
$$   
I_{BS} = \int  \langle {\bf u} ({\bf x},t) \cdot  {\bf u} ({\bf x} + {\bf r},t) \rangle d{\bf r},  \eqno{(5)}
$$   
and Loitsyanskii integrals \cite{dav} ,\cite{my}
$$
I_L =  \int r^2 \langle {\bf u} ({\bf x},t) \cdot  {\bf u} ({\bf x} + {\bf r},t) \rangle d{\bf r}  \eqno{(6)}
$$  
here $<...>$ denotes the ensemble (or spatial) average.   \\

  Conservation of these integrals by the dissipative Navier-Stokes dynamics can be related to the conservation of the momentum and angular momentum, and then to the space homogeneity and isotropy respectively (the Noether’s theorem).\\
  
\begin{figure} \vspace{-1cm}\centering
\epsfig{width=.48\textwidth,file=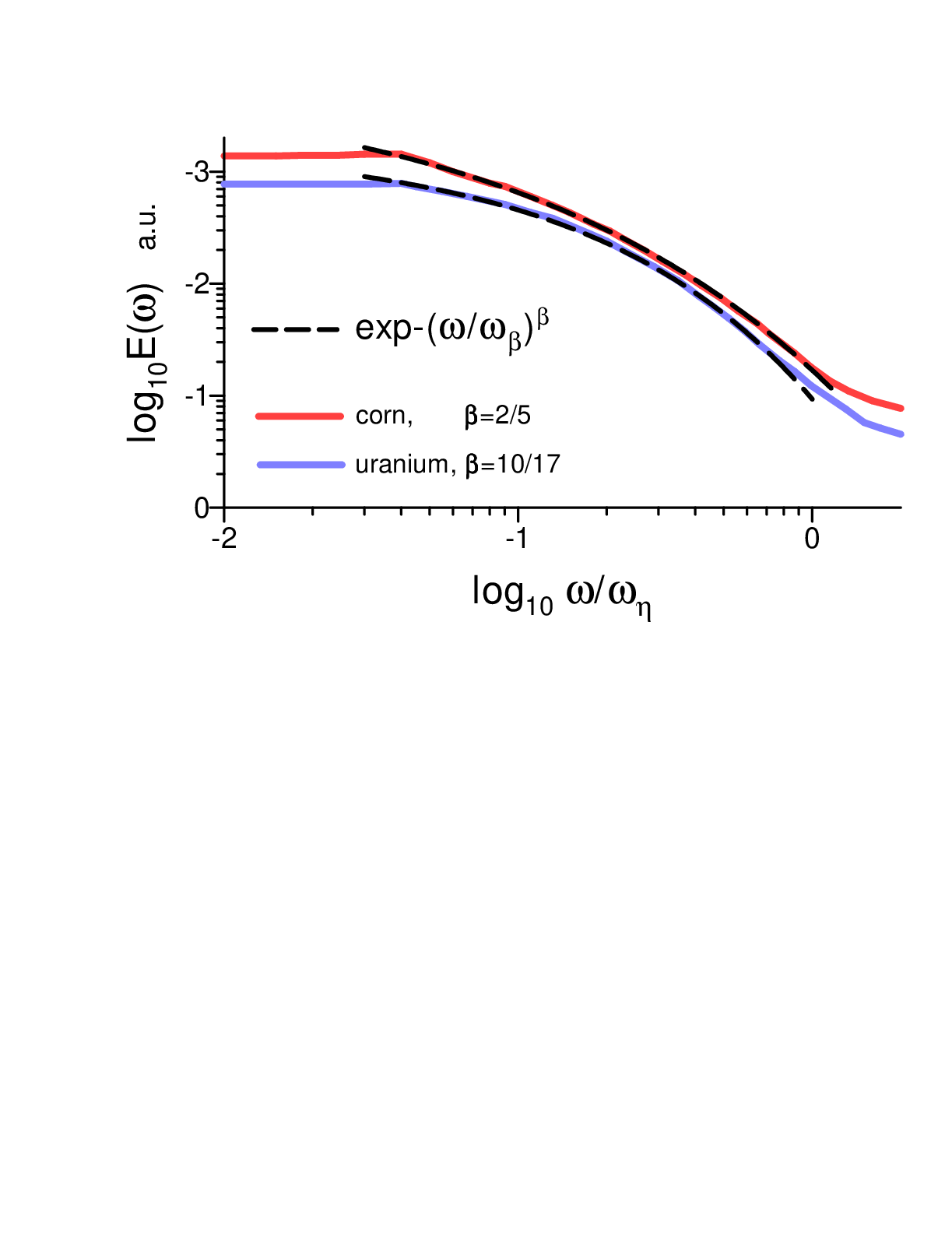} \vspace{-6.1cm}
\caption{The Lagrangian frequency spectra computed for the velocity of fluid surrounding the inertial particles of corn and those of uranium for a particle-laden decaying isotropic homogeneous flow in a cubic box. Unlike the spectra shown in Fig. 3 these spectra were computed in the DNS without gravity. The spectra are vertically shifted for clarity.} 
\end{figure}
\begin{figure} \vspace{-0.45cm}\centering
\epsfig{width=.45\textwidth,file=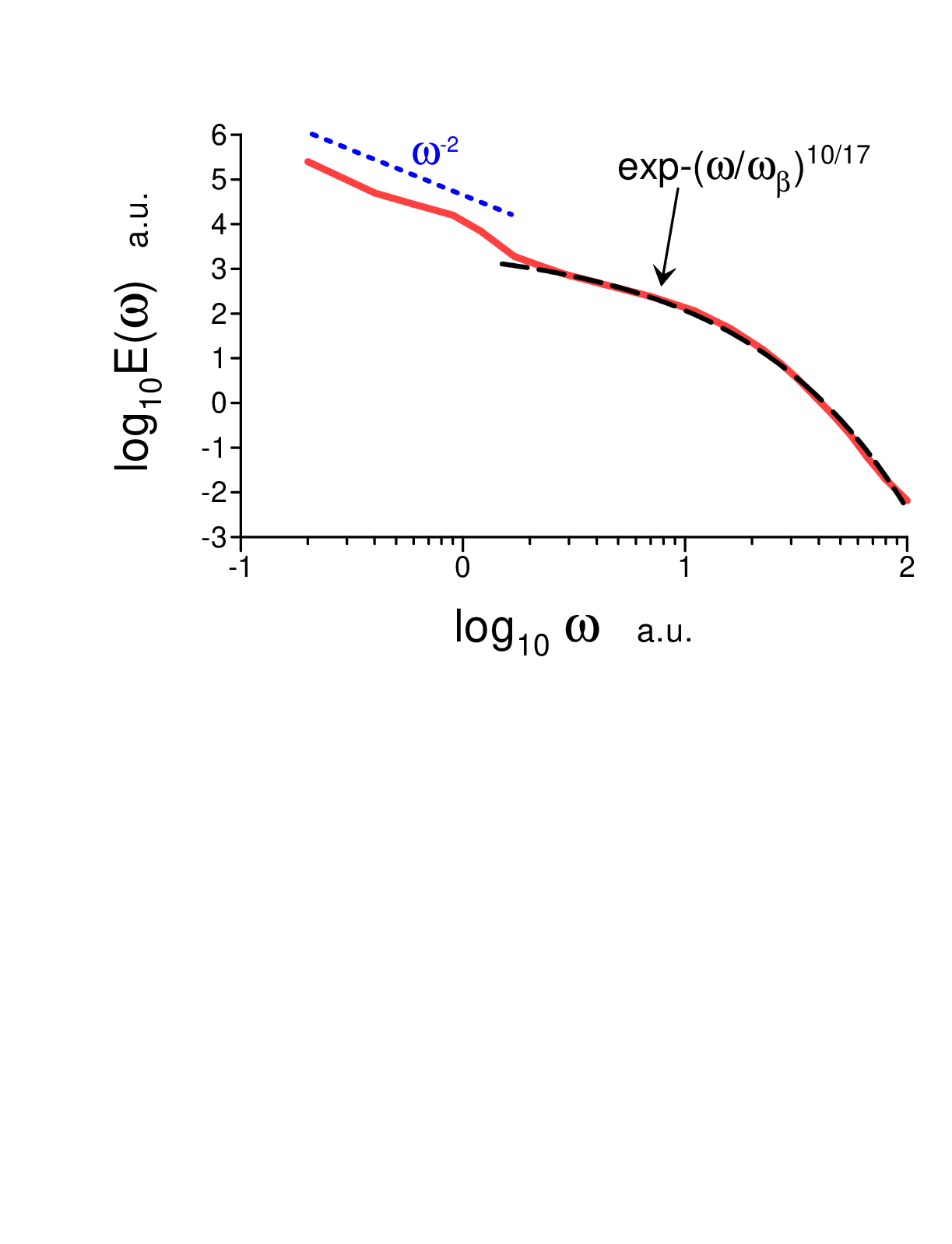} \vspace{-5.1cm}
\caption{The Lagrangian frequency spectra of the particle velocity fluctuations obtained in a DNS of a particle-laden decaying quantum turbulence.} 
\end{figure}

   Let us denote a corresponding dynamic invariant as $I$ and then let us use this invariant to relate the characteristic frequency $\omega_c$ and characteristic velocity $u_c$ using dimensional considerations
 $$
 u_c \propto I^{\delta} \omega_c^{\alpha} \eqno{(7)}
 $$ 

    When the  characteristic velocity $u_c$ is normally distributed \cite{my} the distribution $P(\omega_c)$ can be readily obtained from the relationship Eq. (7).  Comparison of this probability distribution with that provided by the Eq. (4) gives a relationship 
$$
\beta = \frac{2\alpha}{1+2\alpha}  \eqno{(8)}
$$  
between the exponents $\alpha$ and $\beta$.\\

  For the Birkhoff-Saffman invariant Eq. (5) we obtain $\delta =1/5$ and $\alpha =3/5$. Then from the Eq. (8) one obtains
$$
E(\omega) \propto \exp-(\omega/\omega_{\beta})^{6/11}  \eqno{(9)}
$$  

 For the Loitsyanskii invariant Eq. (6) we obtain $\delta =1/7$ and $\alpha = 5/7$. Then from the Eq. (8) one obtains
$$
E(\omega) \propto \exp-(\omega/\omega_{\beta})^{10/17}.  \eqno{(10)}
$$ 

  One can recognize these spectra in the Fig. 1.\\
  
 \section{Lagrangian distributed chaos in more complex flows} 
 
\subsection{Particle-laden flows}  

  For the particle-laden flows the randomization can be more pronounced in the cases when the gravity is not taken into account. Figure 4, for instance, shows the Lagrangian frequency spectra obtained in the same DNS which was mentioned in Section II (Fig. 3) but now without gravity. The spectral data were taken from Fig. 13a of the Ref. \cite{et}. One can see that for the corn particles $\beta  < 1$ (cf. Fig. 3, the concrete value of $\beta = 2/5$ for this case will be discussed in the next sections). For the uranium particles the observed value of $\beta = 10/17$ indicates the distributed chaos dominated by the Loitsyanskii invariant and corresponding spectrum Eq. (10). \\
  
  Figure 5 shows the Lagrangian frequency spectra of the particle velocity fluctuations obtained in a DNS of a particle-laden decaying quantum turbulence. The spectral data were taken from Fig. 10a of the Ref. \cite{uk}. \\
  
  The Gross-Pitaevskii equation was used in this DNS and the particles were modeled as potentials having classical degrees of freedom described by Newtonian dynamics. The well-known Madelung transformation was used to transform the quantum description into a fluid dynamics-like one. It was already shown (see, for instance, Ref. \cite{nab}) that the decay of the incompressible kinetic energy in the decaying superfluid flow shows behavior similar to that observed in decaying classical turbulence (mainly because of the sound emission and mutual friction effects). \\
  
   The dotted straight line is drawn in the Fig. 5 for reference to the Kolmogorov (Lagrangian) scaling $E(\omega) \propto \varepsilon \omega^{-2}$ \cite{ten} for the small frequencies (the hard - nonsmooth turbulence, see Introduction), and the dashed curve indicates the distributed chaos dominated by the Loitsyanskii invariant Eq. (10) for large frequencies (cf Fig. 4).\\ 
   
\begin{figure} \vspace{-0.8cm}\centering
\epsfig{width=.45\textwidth,file=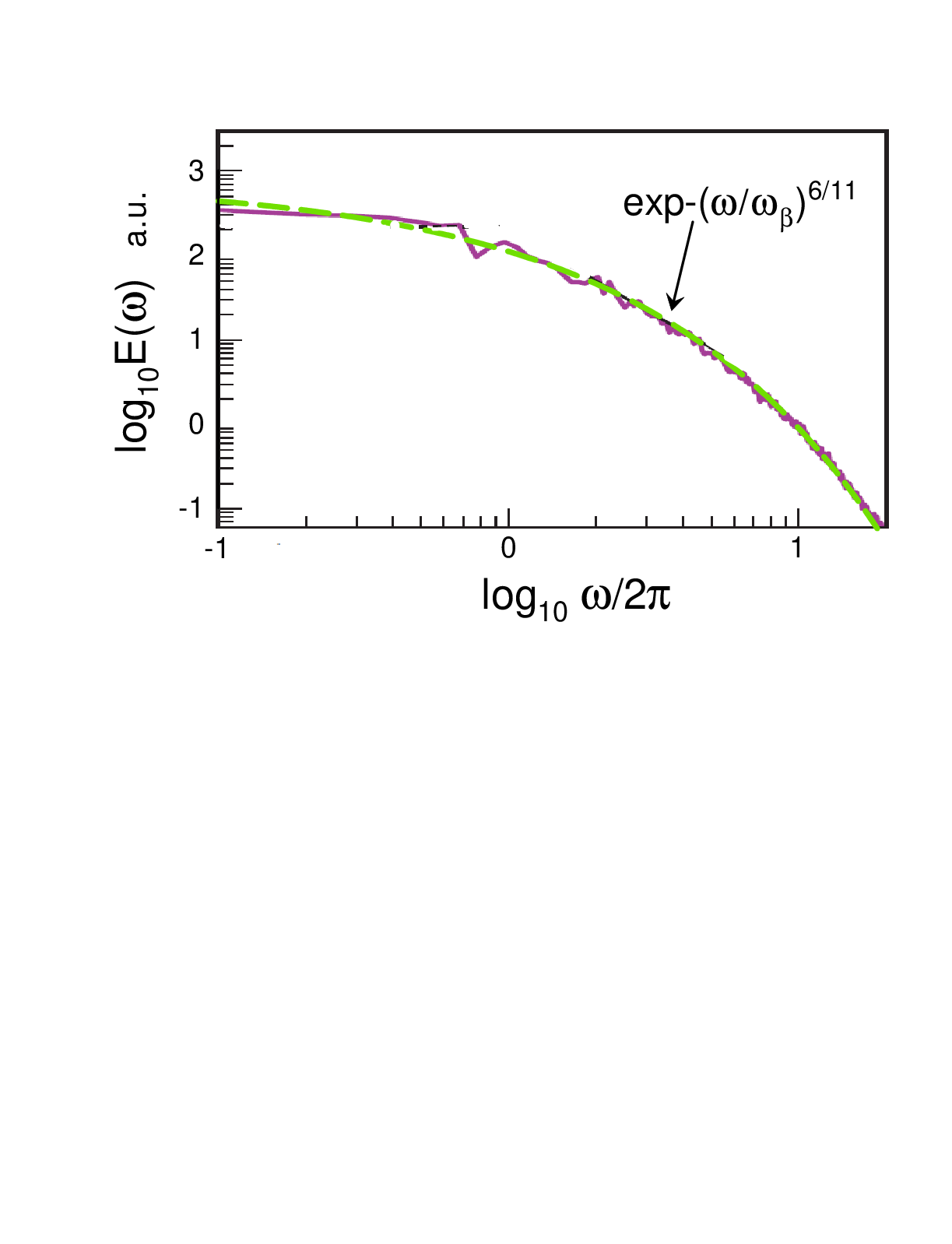} \vspace{-5.3cm}
\caption{The frequency spectra of the $u_z$ component of the velocity field measured in the centre of the cubic domain with the Rayleigh-B\'{e}nard thermal convection. } 
\end{figure}
\begin{figure} \vspace{-0.45cm}\centering
\epsfig{width=.45\textwidth,file=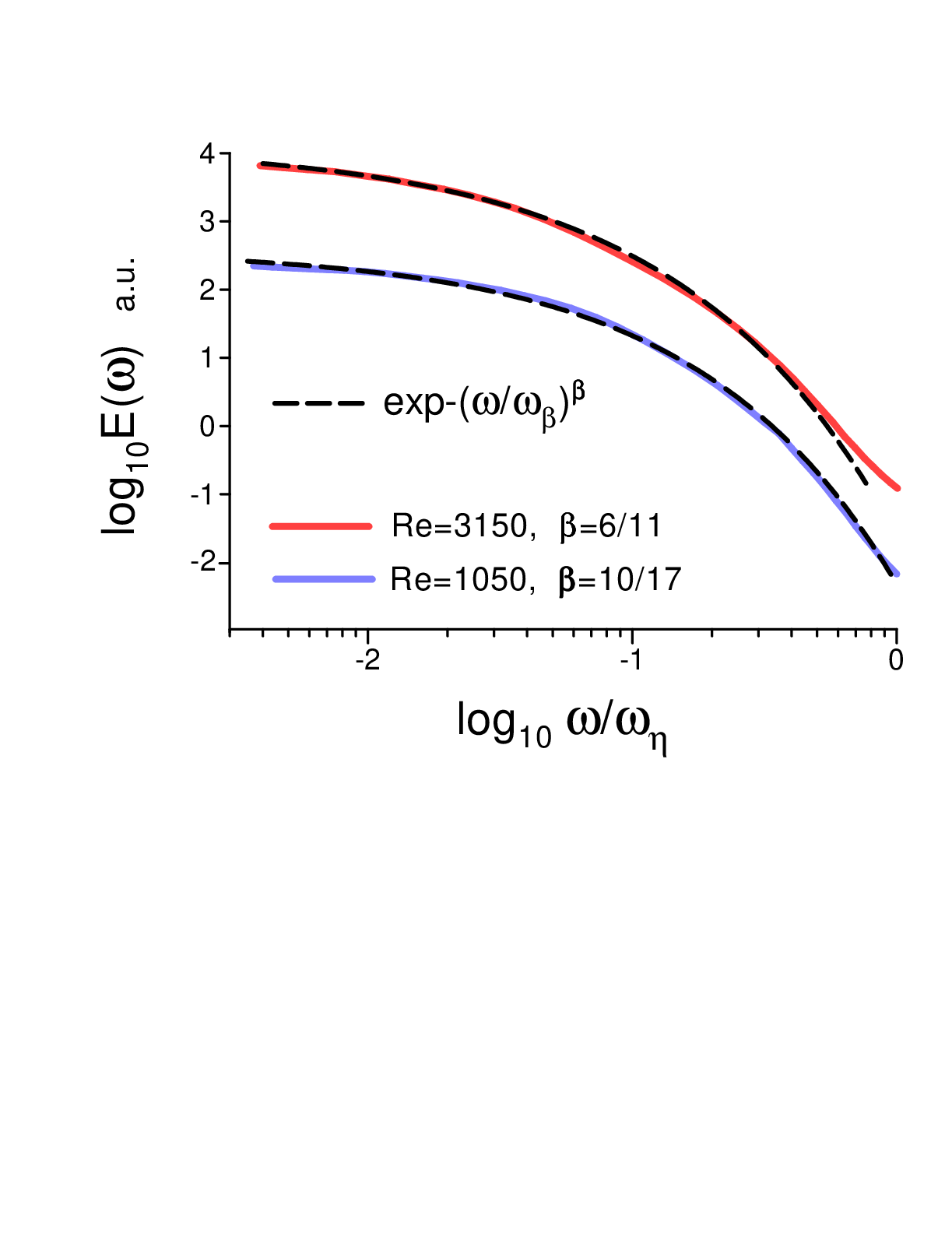} \vspace{-4.3cm}
\caption{The Lagrangian frequency spectra of the fluid velocity computed in the magnetohydrodynamic DNS of a statistically stationary homogeneous isotropic fluid motion for two values of the Reynolds number $Re = 1050,~3150$ (without mean magnetic field).  The spectra are vertically shifted for clarity.} 
\end{figure}

\subsection{Rayleigh-B\'{e}nard convection}  

  Let us now apply the above-obtained results to the Rayleigh-B\'{e}nard thermal convection. The system of equations describing the Rayleigh-B\'{e}nard convection (in the Boussinesq approximation) can be taken as follows
$$
\frac{\partial {\bf u}}{\partial t} + ({\bf u} \cdot \nabla) {\bf u}  =  -\frac{\nabla p}{\rho_0} + \sigma g \theta {\bf e}_z + \nu \nabla^2 {\bf u}   \eqno{(11)}
$$
$$
\frac{\partial \theta}{\partial t} + ({\bf u} \cdot \nabla) \theta  =    \frac{\Delta T}{H} u_z + \kappa \nabla^2 \theta, \eqno{(12)}
$$
$$
\nabla \cdot \bf u =  0 \eqno{(13)}
$$
(see, for instance, Ref. \cite{kcv}). Here $p$ and $\theta$  are the pressure and temperature fluctuation fields respectively ($\theta = T-T_0 (z)$ here $T_0(z)$ is a linear vertical profile of the temperature), ${\bf e}_z$ is the vertical unit vector, $\rho_0$ is the mean density of the liquid, $g$ is the gravity acceleration, $\sigma$ is the thermal expansion coefficient, $\nu$ and  $\kappa$ are the viscosity and thermal diffusivity respectively,  $\Delta T$ is the difference in temperature of the bottom and top boundaries, $H$ is the distance between the bottom and top boundaries. The periodic boundary conditions are usually applied on the side walls,  $\theta =0$ on the bottom and top boundaries as well as the free-slip conditions for velocity. \\ 
  
   For the buoyancy-driven dissipative motion the Birkhoff-Saffman integral can be generalized in the form
 $$
I_{BS} =   \int  \langle {\bf u} ({\bf x},t) \cdot  {\bf u} ({\bf x} + {\bf r},t) -\chi ~ \theta ({\bf x},t)~\theta ({\bf x} + {\bf r},t) \rangle  d{\bf r}  \eqno{(14)}  
$$  
as well as the spectrum Eq. (9) (here $\chi = \alpha g H/\Delta T$).\\

  Figure 6 shows the frequency spectrum of $u_z$ component of the velocity field measured by nine probes located at the centre and vertices of a small cube placed at the centre of a large cubic domain in a DNS of the  Rayleigh-B\'{e}nard convection (for Rayleigh number $Ra=10^8$ and Prandtl number $Pr=1$). The spectral data were taken from Fig. 4 of the Ref. \cite{kv}. Because of the absence of a mean velocity at the centre of the cubic domain a spatial contamination of the temporal spectrum by sweeping effect (see, for instance, \cite{kv} and references therein) is also absent. Therefore, the measured spectrum can be considered as a truly temporal one. \\
  
    The dashed curve in the Fig. 6 indicates spectrum Eq. (9) (the distributed chaos dominated by the generalized Birkhoff-Saffman invariant).\\
    
\subsection{Magnetohydrodynamics} 

The magnetohydrodynamic equations for incompressible isotropic homogeneous case in the Alfv\'enic units are
$$
 \frac{\partial {\bf u}}{\partial t} = - {\bf u} \cdot \nabla {\bf u} 
    -\frac{1}{\rho_0} \nabla {p} - [{\bf b} \times (\nabla \times {\bf b})] + \nu \nabla^2  {\bf u} + {\bf f} \eqno{(15)},
$$
$$
\frac{\partial {\bf b}}{\partial t} = \nabla \times ( {\bf u} \times
    {\bf b}) +\eta \nabla^2 {\bf b}  \eqno{(16)},
$$
$$ 
\nabla \cdot {\bf u}=0, ~~~~~~~~~~~\nabla \cdot {\bf b}=0,  \eqno{(17,18)}
$$
here ${\bf b} = {\bf B}/\sqrt{\mu_0\rho}$  is normalized magnetic field, and $ {\bf f}$ is a forcing function. \\

  It is known that the system (15-18) has the Birkhoff-Saffman and Loitsyanskii invariants as well \cite{dav}, \cite{chan}.\\

    In the paper Ref. \cite{bmg} results of a DNS, using the Eqs. (15-18) in a three-dimensional spatial box with periodic boundary conditions and without a mean magnetic field, were reported. In order to maintain a statistically stationary system the velocity and magnetic field were forced by independent Ornstein-Uhlenbeck processes concentrated in a wavenumber shell around the wavenumber $k_f = 3$. \\

  Figure 7 shows the Lagrangian frequency spectra of the fluid velocity computed in this DNS for two values of the Reynolds number $Re = 1050,~3150$. The spectral data were taken from Fig. 1 of the Ref. \cite{bmg} (three-dimensional case). The dashed curves indicate the spectrum Eq. (9) for $Re = 3150$  and Eq. (10) for $Re =1050$ (distributed chaos dominated by the Birkhoff-Saffman and Loitsyanskii invariants respectively, cf Ref. \cite{ber}) . As it was expected the increase of the Reynolds number results in a decrease of $\beta$, i.e. the randomization becomes stronger with the increase of the Reynolds number. \\

\section{Spontaneous breaking of local reflectional symmetry}
   
 The previous consideration of the distributed chaos was concentrated on isotropic homogeneous cases with global reflectional symmetry. The global reflection symmetry results in zero mean (global) helicity. 
 
   However, spontaneous breaking of the {\it local} reflectional symmetry can result in the appearance of the nonzero point-wise helicity in this case. This phenomenon is an intrinsic property of the chaotic/turbulent flows (see, for instance, Refs. \cite{bkt},\cite{kerr},\cite{hk}). 
   
   The appearance of the vorticity blobs having nonzero helicity and moving with the fluid should accompany this phenomenon (see, for instance, Refs. \cite{moff1}-\cite{bt} and references therein). These vorticity blobs having high relative helicity are a generic property of the chaotic/turbulent flows \cite{moff2}. The low viscous dissipation over these blobs would result in adiabatic invariance of their associated helicity \cite{moff1},\cite{mt}. \\
 
   Since the global helicity should be identically equal to zero at this phenomenon the localized positive and negative helicities associated with the blobs have to be canceled at the global average.\\

   The helicity associated with a vorticity blob is
$$
H_j = \int_{V_j} h({\bf r},t) ~ d{\bf r}.  \eqno{(19)}
$$
here $h({\bf r},t) = {\bf u} \cdot {\boldsymbol \omega}$ is the helicity distribution, $V_j$ is the spatial volume of the j-blob, and ${\boldsymbol \omega} ({\bf r},t)= \nabla \times {\bf u}  ({\bf r},t)$ is the vorticity field.\\

   Moments of the helicity distribution  $h({\bf r},t) = {\bf u} \cdot {\boldsymbol \omega}$ are defined as \cite{lt} ,\cite{mt}
$$
{\rm I_n} = \lim_{V \rightarrow  \infty} \frac{1}{V} \sum_j H_{j}^n  \eqno{(20)}
$$
where $V$ is the blobs' total volume.\\

   Let $H_j^{-}$ be the total helicity associated with the blobs with negative helicity and $H_j^{+}$ be the total helicity associated with the blobs with positive helicity. Then the sign-defined moments of the helicity distribution are
$$
{\rm I_n^{\pm}} = \lim_{V \rightarrow  \infty} \frac{1}{V} \sum_j [H_{j}^{\pm}]^n  \eqno{(21)}
$$ 
The summation in Eq. (21) is performed for the blobs with negative (or positive) $H_{j}^{\pm}$ only.  \\

  For the odd values of $n$ we have ${\rm I_n} = {\rm I_n^{+}} + {\rm I_n^{-}} =0$, due to the global reflectional symmetry. Hence, for the odd values of $n$ we obtain ${\rm I_n^{+}} = - {\rm I_n^{-}}$.\\
 
  The blobs with high associated helicity should provide the main contribution to the moments with high values of $n$ \cite{bt} (though, the value $n=2$ can be often considered sufficiently high for the strongly intermittent chaotic/turbulent cases). That provides additional support for the use of the helicity distribution moments ${\rm I_n}$ and ${\rm I_n^{\pm}}$ as the adiabatic invariants.\\

\section{Lagrangian distributed chaos and moments of helicity distribution}

    If a  flow is dominated by the adiabatic invariant $I_n$ for even moments or by the adiabatic invariant $I_n^{\pm}$ for odd moments we can use the dimensional considerations 
$$
 u_c \propto |I_n|^{1/(4n-3)}~ \omega_c^{\alpha_n}    \eqno{(21)}
$$ 
 or
$$
 u_c \propto |I_n^{\pm}|^{1/(4n-3)}~ \omega_c^{\alpha_n}    \eqno{(22)}
$$   
 for the even or for the odd moments respectively. 
 For the both cases
$$
\alpha_n = \frac{2n-3}{4n-3}  \eqno{(23)}
$$  

 Then, from Eq. (8) we obtain
$$
 \beta_n = \frac{2(2n-3)}{(8n-9)}   \eqno{(24)}  
 $$
 
In particular, for the two end cases: \\
 
 $n \gg 1$ - weakly randomized helical distributed chaos:
$$
E(k) \propto \exp-(k/k_{\beta})^{1/2},  \eqno{(25)}
$$ 

\begin{figure} \vspace{-0.5cm}\centering
\epsfig{width=.46\textwidth,file=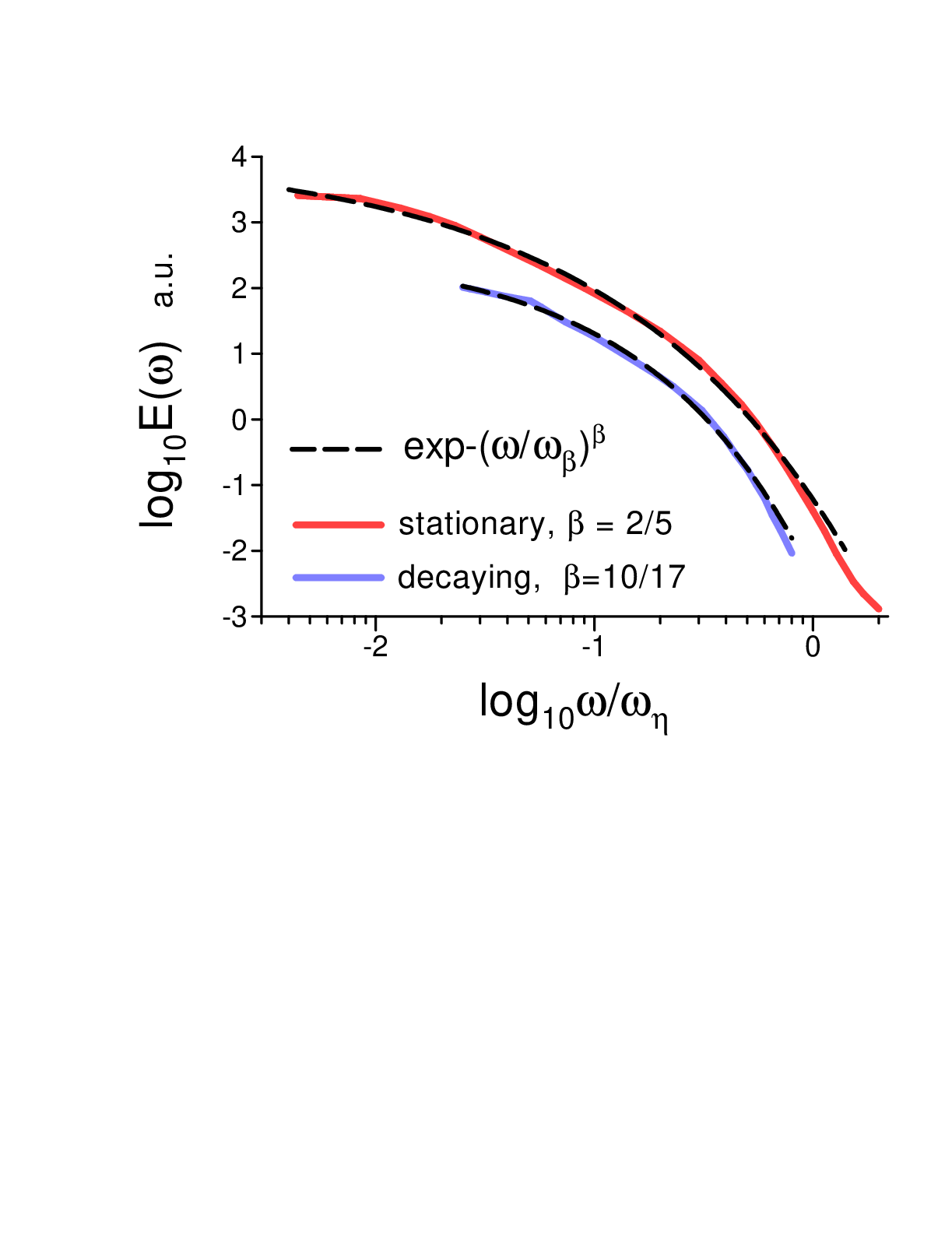} \vspace{-4.55cm}
\caption{Lagrangian frequency spectrum of velocity in isotropic homogeneous fluid motion: statistically stationary (upper curve, $Re_{\lambda} = 240$) and decaying (lower curve). The spectra are vertically shifted for clarity. } 
\end{figure}
\begin{figure} \vspace{-0.5cm}\centering
\epsfig{width=.46\textwidth,file=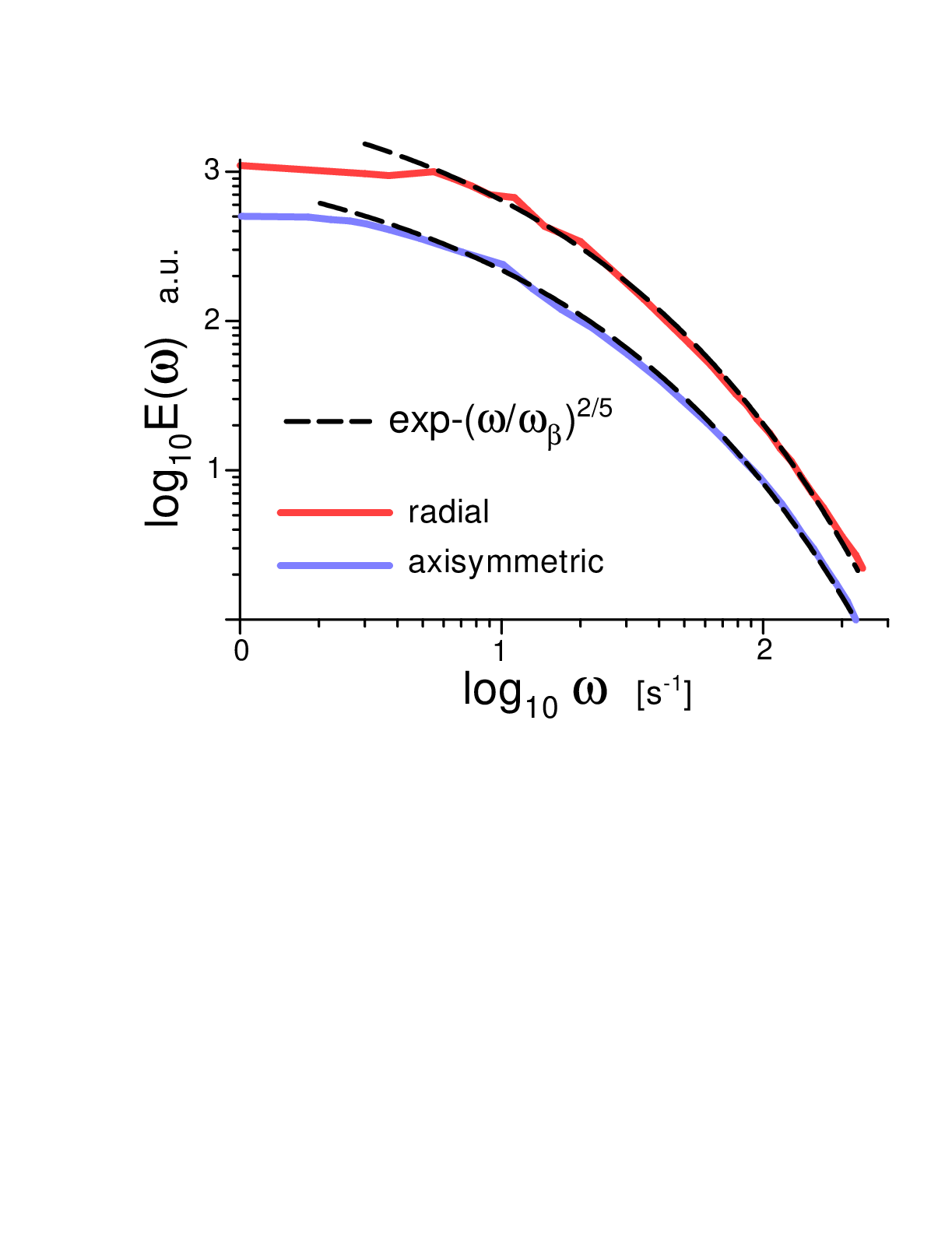} \vspace{-4.55cm}
\caption{Lagrangian frequency spectra of velocity along a passive tracer trajectory for $Re_{\lambda} = 124$ in a laboratory experiment \cite{berg}. The spectra are vertically shifted for clarity.} 
\end{figure}

and for $n=2$ (the Levich-Tsinober invariant \cite{lt}) - strongly randomized helical distributed chaos:
$$
E(k) \propto \exp-(k/k_{\beta})^{2/7}  \eqno{(26)}
$$   

 For a moderate randomization $n=3$
 $$
 E(k) \propto \exp-(k/k_{\beta})^{2/5}  \eqno{(27)}
$$

   One can recognize spectrum Eq. (27) in the Fig. 1 for $Re_{\lambda} = 234$. This figure shows the Lagrangian frequency spectra obtained for statistically stationary isotropic and homogeneous chaotic fluid motion computed in a DNS \cite{pk} (see a description of the DNS in the Introduction). \\
   
   An analogous spectrum was obtained in a DNS of statistically stationary isotropic and homogeneous chaotic motion of an incompressible fluid reported in the paper Ref. \cite{luc} for $Re_{\lambda} = 240$. The spectral data were taken from Fig. 3 of the Ref. \cite{luc}. In this DNS the statistically stationary chaotic motion was maintained using Lundgren’s linear forcing \cite{lun} by applying a force proportional to the velocity. \\
   
   Figure 8 shows the spectra computed in this DNS for two types of isotropic homogeneous motion: statistically stationary (upper curve) and freely decaying (lower curve). The spectral data were taken from Figs. 1 and 6 of the Ref. \cite{luc} respectively. The dashed curves indicate the spectra Eq. (27) for statistically stationary and Eq. (10) for decaying cases (cf. Fig. 1 and comments to the Fig. 1 in the Introduction). 
   
\section{Lagrangian helical distributed chaos in more complex flows}   
   
   One can also recognize spectrum Eq. (27) in Fig. 4 for the particle-laden flow (helical distributed chaos for the corn particles). If we compare Fig. 4 with Fig. 3 for the corn particles we can see that the deterministic chaos ($\beta = 1$ for the case with gravitation Fig. 3) has been replaced by the helical distributed chaos ($\beta = 2/5$ for the case without gravitation Fig. 4), i.e. a randomization takes place for the case without gravitation. \\ 
   
   Figure 9 shows the Lagrangian frequency spectra of velocity along a passive tracer trajectory for $Re_{\lambda} = 124$ in a laboratory experiment \cite{berg}. The spectral data were taken from Fig. 5b of the Ref. \cite{berg}.\\
   
   In this experiment, the flow of water was generated by eight propellers located in the corners of a rectangular tank ($32\times 32 \times 50cm^3$), and the axis of rotation for all propellers was in vertical direction. The rotational direction of the propellers was changed in fixed intervals of time to suppress the existing mean flow. The measurements were made by the PTV method using neutrally buoyant illuminated particles (with a very low Stokes number) inside a spatial ball with a radius of 50mm centered in the center of the experimental tank. Inside this ball the illuminated particles were uniformly distributed. \\
   
   Figure 9 shows the spectra of the radial and axisymmetric (vertical) velocity components. The dashed curves indicate the spectrum Eq. (27) for both components.\\
   
   Results of another laboratory experiment with a fan-stirred quasi-isotropic homogeneous flow were reported in paper Ref. \cite{blm}. The flow was generated in a large-scale spherical vessel by four rotating fans placed near the wall and was quasi-isotropic and homogeneous in a central spatial ball inside the experimental vessel. The velocity characteristics were measured with a Particle Image Velocimetry (PIV) technique for $Re_{\lambda} = 220.2; 385$; and $555.4$.\\

\begin{figure} \vspace{-1.1cm}\centering \hspace{-1.8cm}
\epsfig{width=.52\textwidth,file=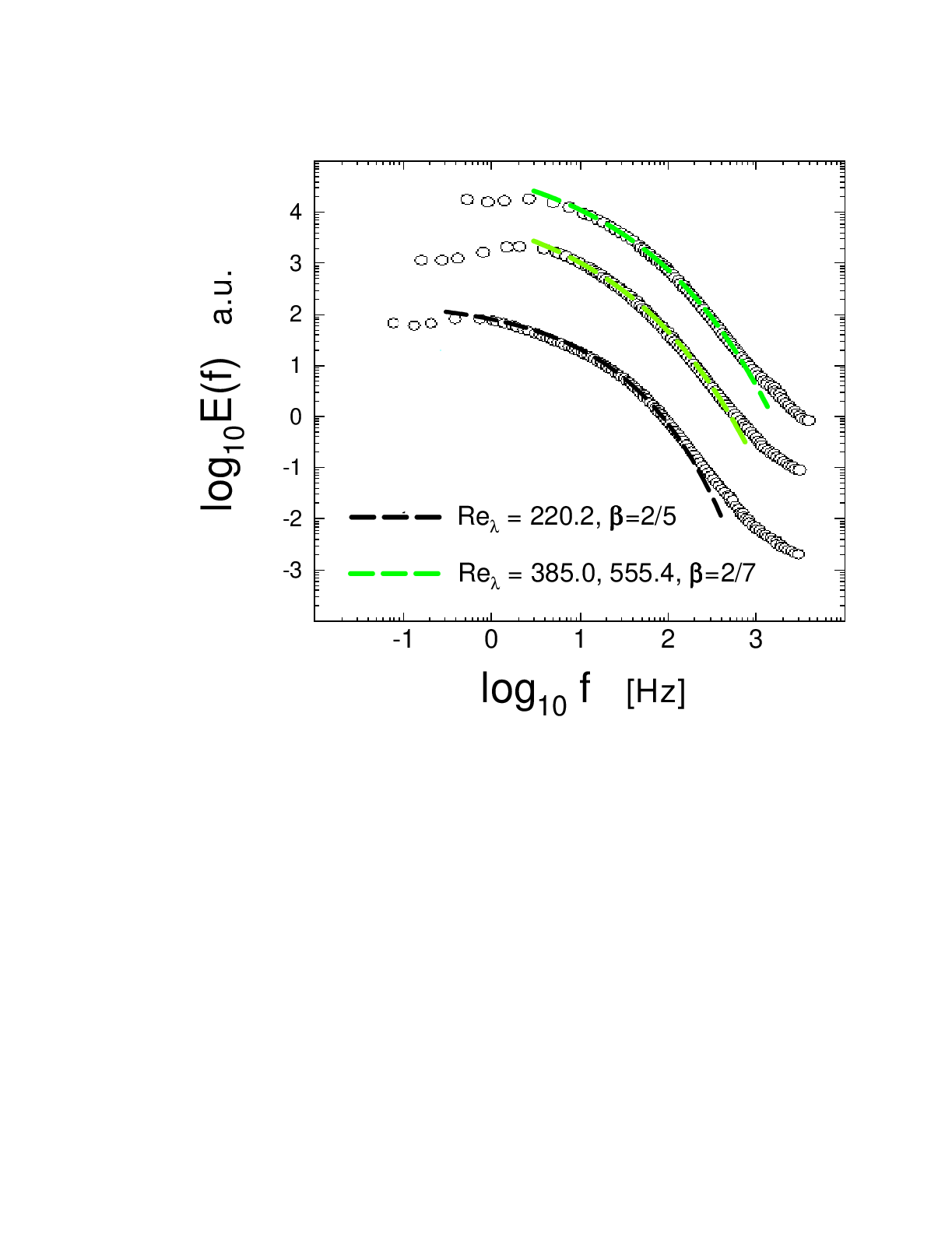} \vspace{-5.2cm}
\caption{Frequency spectra of velocity measured with a PIV technique for $Re_{\lambda} = 220.2; 385$ and $555.4$ in a laboratory experiment \cite{blm}. } 
\end{figure}
\begin{figure} \vspace{-0.5cm}\centering 
\epsfig{width=.45\textwidth,file=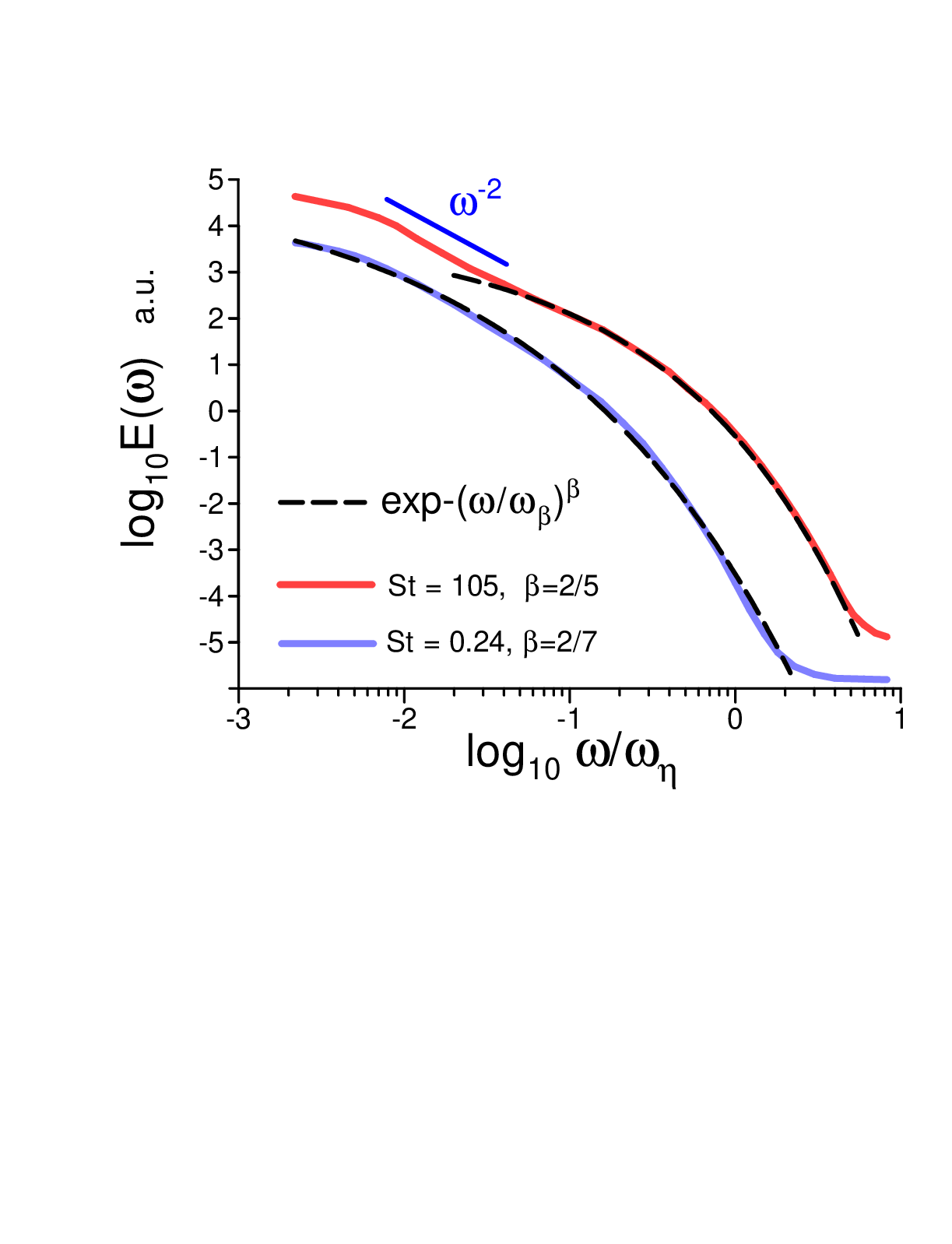} \vspace{-4cm}
\caption{Lagrangian frequency spectra of the fluid velocity along the particle trajectory for a heavy-particle-laden isotropic homogeneous fluid motion  ($Re_{\lambda} = 400$ and only the Stokes drag force was taken into account in the DNS.)} 
\end{figure}
\begin{figure} \vspace{-0.8cm}\centering 
\epsfig{width=.48\textwidth,file=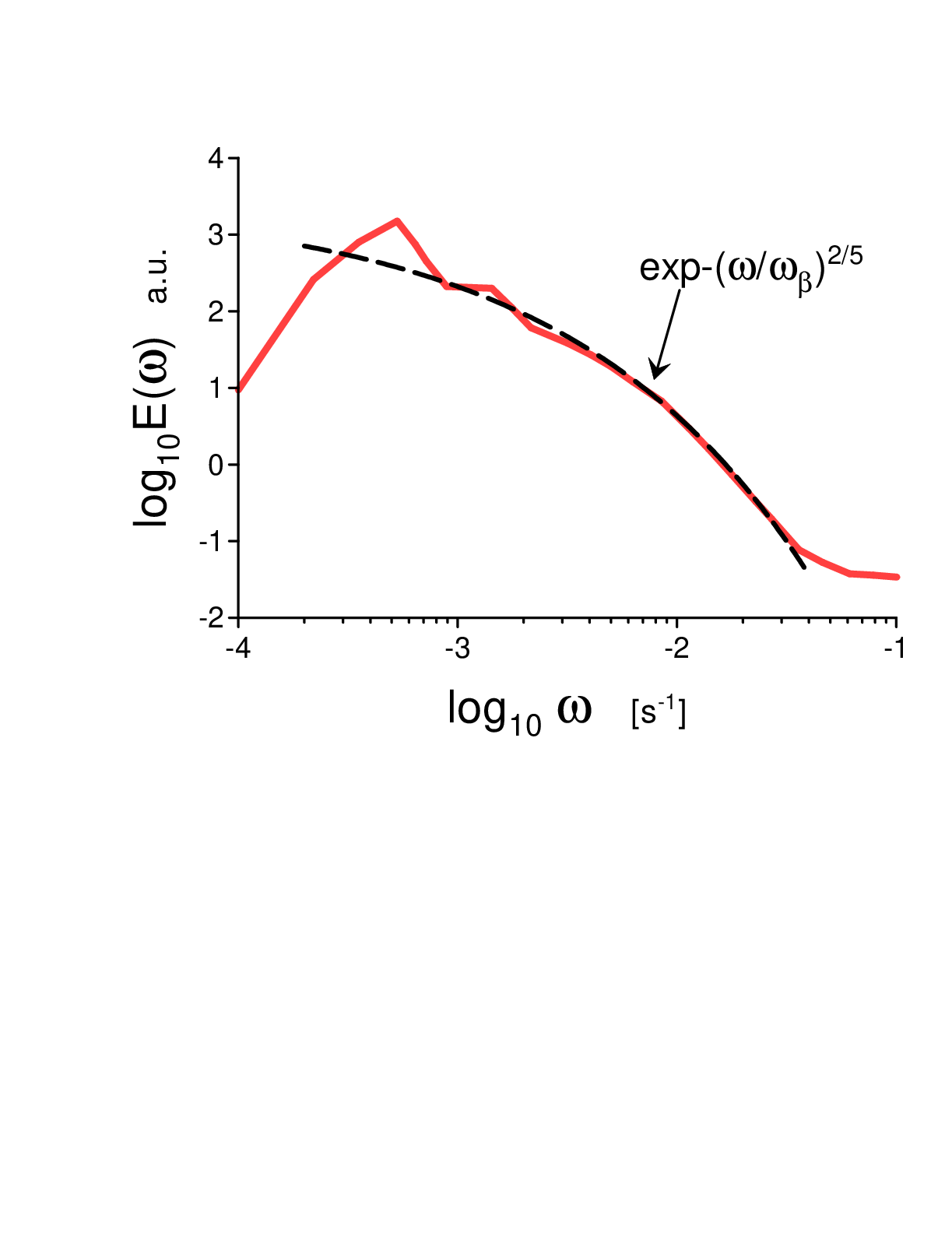} \vspace{-4.95cm}
\caption{Lagrangian frequency spectra of vertical velocity obtained using the measurements made by a single Lagrangian float in the nighttime convective boundary layer.} 
\end{figure}

 Figure 10 shows the corresponding spectra (the spectral data were taken from Fig. 11 of the Ref. \cite{blm}. The dashed curves indicate the spectral laws Eq. {27} for $Re_{\lambda} = 220.2$ and Eq. (26) for $Re_{\lambda} = 385$ and $555.4$. As it was expected the value of the dimensionless parameter $\beta =2/5$ for $Re_{\lambda} = 220.2$ is larger then that for $Re_{\lambda} = 385$ and $555.4$ ($\beta =2/7$). I.e. the randomization of the chaotic flow is lesser for the smaller $Re_{\lambda}$ (see Introduction). \\
   
  Figure 11 shows the Lagrangian frequency spectra of the fluid velocity along the particle trajectory for a heavy-particle-laden isotropic homogeneous fluid motion ($Re_{\lambda} = 400$ and only the Stokes drag force was taken into account ($St$ is the Stokes number). The DNS obtained spectral data were taken from Fig. 6a of the Ref. \cite{zlz} (see also Refs. \cite{bec},\cite{lan}). The dashed curves indicate the spectra Eq. (26) for $St=0.24$ and Eq. (27) for $St=105$. The straight line is drawn for reference to the Kolmogorov (Lagrangian) scaling $E(\omega) \propto \varepsilon \omega^{-2}$ \cite{ten} (the hard - nonsmooth turbulence, see Introduction).\\
  
  Interesting examples of the Lagrangian helical distributed chaos were obtained in the oceanic subsurface observations with the Lagrangian floats \cite{afod}, \cite{al}. \\
  
 In the first example, a subsurface mixed layer (80 m deep) was subject to nighttime radiational cooling that produced a buoyancy-driven convection. Figure 12 shows the Lagrangian spectrum of vertical velocity obtained using the measurements made by a single Lagrangian float in the nighttime convective boundary layer. The spectral data were taken from Fig. 8 of the Ref. \cite{afod}. The dashed curve indicates the spectrum Eq. (27) (cf Fig. 2). \\
 
  In the second example, a Lagrangian frequency spectrum was computed from a set of Lagrangian float trajectories in Knight Inlet where chaotic/turbulent mixing coexists with energetic internal waves. The measurements were made by two floats simultaneously launched upstream of the Knight Inlet sill in the presence of a strong flood tide. \\
  
  Figure 13 shows the Lagrangian frequency spectra of the vertical velocity obtained from the float trajectories traversing the sill (the spectral data were taken from Fig. 6b of the Ref. \cite{al}). The dashed curve indicates the spectrum Eq. (26).\\
  
\begin{figure} \vspace{-1.2cm}\centering 
\epsfig{width=.48\textwidth,file=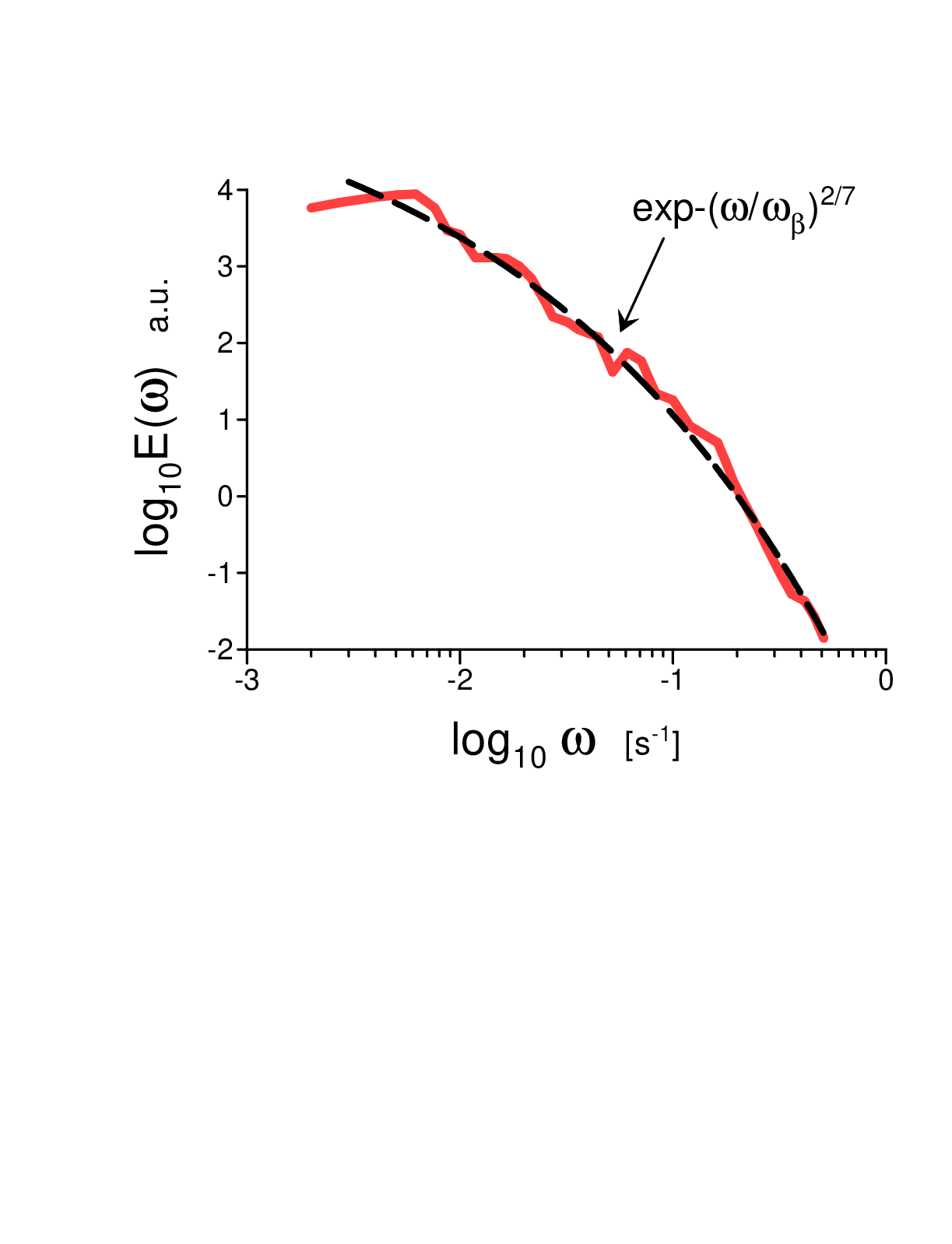} \vspace{-4.8cm}
\caption{Lagrangian frequency spectra of vertical velocity obtained using the measurements made by two Lagrangian floats that traverse an inlet sill on strong flood tide.} 
\end{figure}
\begin{figure} \vspace{-0.5cm}\centering 
\epsfig{width=.48\textwidth,file=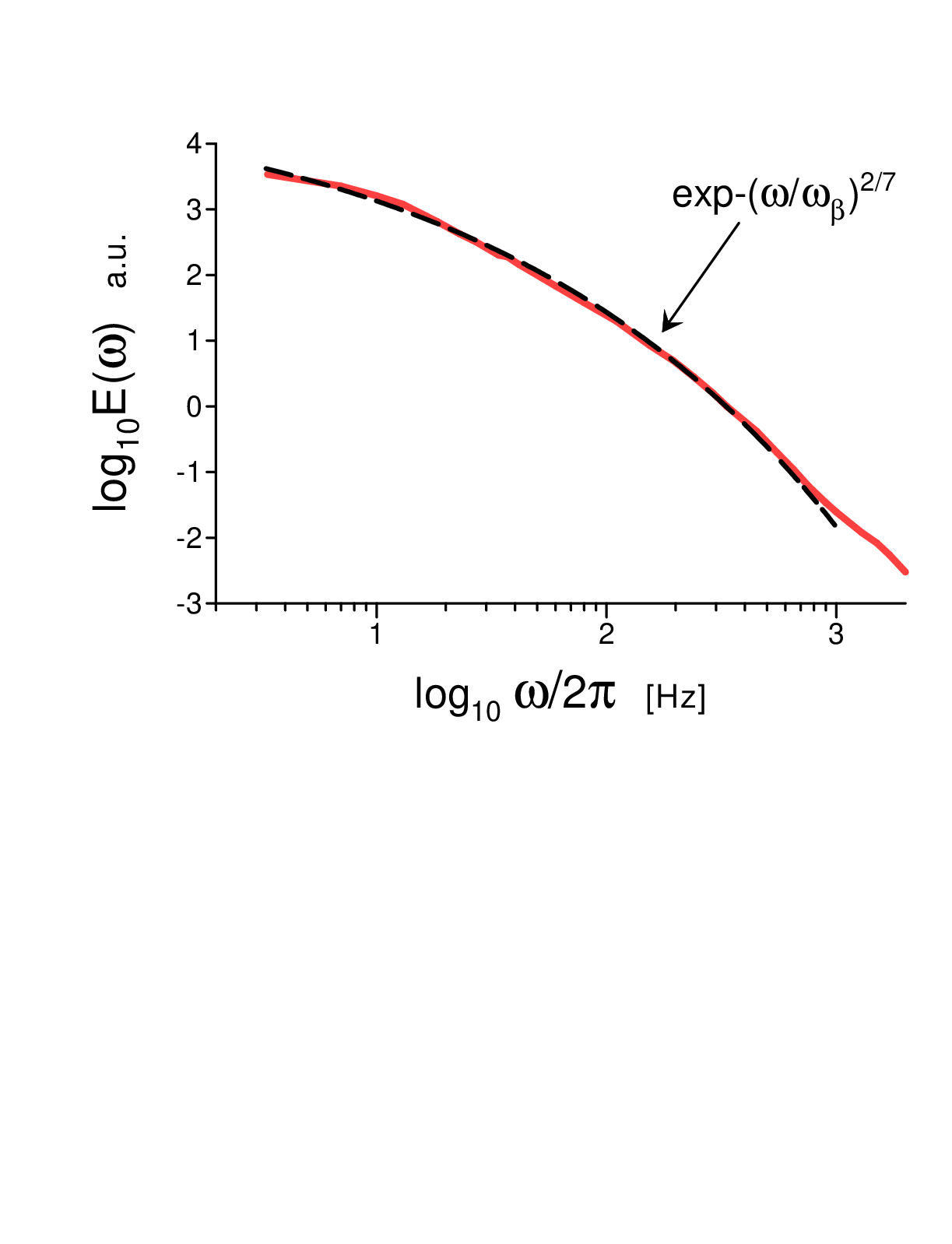} \vspace{-5.2cm}
\caption{Lagrangian frequency spectra of the fluid velocity obtained in a von Karman-like laboratory `swirling' experiment (with two counterrotating disks) for $Re_{\lambda} = 740$.}
\end{figure}
   
   In a laboratory experiment, described in paper Ref. \cite{mo}, a swirling flow was created at high $Re_{\lambda} = 740$ using two counterrotating disks (fitted with blades) located at the opposite sides of a cylindrical vessel (a von Karman type of experimental setting). In the center of the vessel the velocity fluctuation can be considered quasi-isotropic and homogeneous. Tracking of the tracer particles was performed using an acoustic technique.\\ 
   
   Figure 14 shows the power spectrum of the measured Lagrangian velocity. The spectral data were taken from Fig. 2b of the Ref. \cite{mo}. The dashed curve indicates the spectrum Eq. (26).\\

 \section{Conclusions}
   
 The notion of distributed chaos can be used to quantify the randomization of the Lagrangian chaos. The main dimensionless parameter of the distributed chaos - $\beta$ Eq. (1), has been applied for this purpose. Relevant dissipative dynamical invariants - the Birkhoff-Saffman and Loitsyanskii integrals, as well as helical dynamical invariants (moments of helical distribution) related to the spontaneous breaking of local reflectional symmetry, have been used to find the corresponding values of $\beta$ for different flows: isotropic homogeneous fluid motions, buoyancy-driven, particle-laden, magnetohydrodynamic, chaotic/turbulent motions of quantum fluids, and some subsurface oceanic flows (mixing). \\
 
   Using this approach it is shown, in particular, that an increase of $Re_{\lambda}$ (i.e. intensification of the fluid motion) results in an increase of randomization of the Lagrangian chaos (see, for instance, Fig. 1, Fig. 7, Fig. 10 and Fig. 14). For the particle-laden flows gravitation suppresses the randomization (cf Fig. 3 with Fig. 4). In some cases the Kolmogorov scaling (in the Lagrangian interpretation $E(\omega) \propto \varepsilon \omega^{-2}$) can coexist with the Lagrangian distributed chaos (in different ranges of scales, see Fig. 5 and Fig. 11). 
 
\section{Acknowledgments}
 
 I thank H.K. Moffatt for drawing my attention to paper Ref. \cite{moff2}, and E. Levich for discussions.


\begin{thebibliography}{99}
\bibitem{aref} H. Aref, J. Fluid Mech., {\bf 143}, 1 (1984)
\bibitem{fm} U. Frisch and R. Morf, Phys. Rev., {\bf 23}, 2673 (1981)
\bibitem{oh} N. Ohtomo, K. Tokiwano, Y. Tanaka et. al., J. Phys. Soc.
Jpn., {\bf 64}, 1104 (1995)
\bibitem{mm1} J. E. Maggs and G. J. Morales, Phys. Rev. Lett., {\bf 107},185003 (2011) 
\bibitem{mm2} J. E. Maggs and G. J. Morales, Phys. Rev. E {\bf 86}, 015401(R) (2012)
\bibitem{wu} X-Z. Wu, L. Kadanoff, A. Libchaber, and M. Sano, Phys.
Rev. Lett., {\bf 64}, 2140 (1990)
\bibitem{pk} P.K Yeung, J. Fluid Mech., {\bf 427}, 241 (2001)
\bibitem{kds} S. Khurshid, D.A. Donzis and K.R. Sreenivasan, Phys. Rev. Fluids, {\bf 3}, 082601(R) (2018)
\bibitem{rupolo} V. Rupolo, J. Phys. Oceanogr., {\bf 37}, 1584 (2007)
\bibitem{et} S. Elghobashi, G.C. Truesdell, J. Fluid. Mech.,  {\bf 242}, 655 (1992)
\bibitem{jon} D.C. Johnston, Phys. Rev. B, {\bf 74}, 184430 (2006)
\bibitem{bir} G. Birkhoff, Commun. Pure Appl. Math., {\bf 7}, 19 (1954)
\bibitem{saf} P. G. Saffman, J. Fluid. Mech., {\bf 27}, 551 (1967)
\bibitem{dav} P.A. Davidson, J. Fluid Mech., {\bf 663}, 268 (2010)
\bibitem{my} A. S. Monin, and A. M. Yaglom, Statistical Fluid Mechanics, Vol. II: Mechanics of Turbulence (Dover Pub. NY, 2007)
\bibitem{uk} U. Giuriato, and G. Krstulovic, Phys. Rev. Fluids, {\bf 5}, 054608 (2020)
\bibitem{ten} H. Tennekes, J. Fluid Mech., {\bf 67}, 561 (1975)
\bibitem{nab} C. Nore, M. Abid, and M.E. Brachet, Phys. Fluids, {\bf 9}, 2644 (1997)
\bibitem{kcv} A. Kumar, A.G. Chatterjee and M.K. Verma, Phys. Rev. E, {\bf 90}, 023016 (2014)
\bibitem{kv} A. Kumar, and M.K. Verma, R. Soc. open sci., {\bf 5}, 172152 (2018)
\bibitem{chan} S. Chandrasekhar, Proc. R. Soc. London, Ser. A, {\bf 204}, 435 (1951).
\bibitem{bmg} A. Busse, W.C. Müller, and G. Gogoberidze, Phys. Rev. Lett., {\bf 105}, 235005 (2010)
\bibitem{ber} A Bershadskii, Res. Notes AAS, {\bf 4}, 10 (2020)
\bibitem{bkt} A. Bershadskii, E. Kit, A. Tsinober, Proc. R. Soc. Lond. A, {\bf 441}, 147 (1993)
\bibitem{kerr} R.M. Kerr,  In: Elementary Vortices and Coherent Structures, Proceedings of the IUTAM Symposium Kyoto, 1-8 (2004)
\bibitem{hk} D.D. Holm, R.M. Kerr, Physics of Fluids, {\bf 19}, 025101 (2007)
\bibitem{moff1} H.K. Moffatt, J . Fluid Mech., {\bf 35}, 117 (1969)
\bibitem{moff2} H.K. Moffatt, J . Fluid Mech., {\bf 159}, 359 (1985)
\bibitem{lt} E. Levich and A. Tsinober, Phys. Lett. A {\bf 93}, 293 (1983)
\bibitem{mt} H.K. Moffatt and A. Tsinober, Annu. Rev. Fluid Mech., {\bf 24}, 281 (1992)
\bibitem{bt} A. Bershadskii and A. Tsinober,  Phys. Rev. E, {\bf 48}, 282 (1993)
\bibitem{luc} F. Lucci, V.S. L’vov , A. Ferrante, M. Rosso , and S. Elghobashi, Theor. Comput. Fluid Dyn., {\bf 28}, 197 (2014)
\bibitem{lun} T.S. Lundgren, Annual Research Briefs, Center for Turbulence Research, pp. 461–473, Stanford (2003)
\bibitem{berg} J. Berg, 	arXiv:physics/0610155 (2006)
\bibitem{blm} D. Bradley, M. Lawes, and M.E. Morsy, Journal of Turbulence, {\bf 20}, 195 (2019)
\bibitem{zlz} Z. Zhang, D. Legendre, and R. Zamansky, J. Fluid Mech., {\bf 921}, A4 (2021)
\bibitem{bec} J Bec, L Biferale, M Cencini, AS Lanotte, F. Toschi, J. Fluid Mech., {\bf 646}, 527 (2010)
\bibitem{lan} A. Lanotte, E. Calzavarini, T. Federico, B. Jeremie, B. Luca, and C. Massimo, Heavy particles
in turbulent flows. International CFD Database (2011)
\bibitem{afod} E.A. D'Asaro, D.M. Farmer, J.T. Osse, and G.T. Dairiki, J. Atmos. Oceanic Technol., {\bf 13}, 1230 (1996)
\bibitem{al}  E.A. D'Asaro and R-C. Lien, J. Phys. Oceanogr., {\bf 30}, 641 (2000)
\bibitem{mo} N. Mordant, J. Delour, E. L\'{e}veque, O. Michel, A. Arn\'{e}odo, and J.-F. Pinton, J. Stat.Phys. {\bf 113}, 701 (2003)
\end{thebibliography}
\end{document}